\documentclass[twocolumns]{aa} 
\usepackage{graphicx}
\usepackage{txfonts}
\usepackage{amsmath,amsfonts,amssymb} 
\usepackage{natbib}
\usepackage{here}
\usepackage{comment}
\bibpunct{(}{)}{;}{a}{}{,}
\usepackage[colorlinks=true,urlcolor=blue,linkcolor=blue,citecolor=blue]{hyperref}
\usepackage[normalem]{ulem}
\usepackage{color}

\newcommand\rh{r$_h$}

\begin{document} 

\title{Activity and composition of the hyperactive comet 46P/Wirtanen during its  close approach in 2018}

\authorrunning{Moulane et al. 2022}
\titlerunning{Comet 46P/Wirtanen}
        
\author{
Y. Moulane\inst{1}, 
E. Jehin\inst{2}, 
J. Manfroid\inst{2},
D. Hutsemékers\inst{2},
C. Opitom\inst{3},
Y. Shinnaka\inst{4},
D. Bodewits\inst{1},
Z. Benkhaldoun\inst{5},
A. Jabiri\inst{5},
S. Hmiddouch\inst{2,5}, 
M. Vander Donckt\inst{2},
F. J. Pozuelos\inst{2,6},\and
B. Yang\inst{7}
     }
           
\institute{
Physics Department, Leach Science Center, Auburn University, AL 36832, USA\\
$^\dagger$\email{\color{blue}ymoulane@auburn.edu}
\and
Space sciences, Technologies \& Astrophysics Research (STAR) Institute, University of Liège, Liège, Belgium
\and
Institute for Astronomy, University of Edinburgh, Royal Observatory, Edinburgh EH9 3HJ, UK \and
Koyama Astronomical Observatory, Kyoto Sangyo University, Motoyama, Kamigamo, Kita-ku, Kyoto 603-8555, Japan 
\and
Oukaimeden Observatory, High Energy Physics and Astrophysics Laboratory, Cadi Ayyad University, Marrakech, Morocco
\and
Astrobiology Research Unit, Université de Liège, Allée du 6 Août 19C, B-4000 Liège, Belgium
\and
Núcleo de Astronomía, Facultad de Ingeniería, Universidad Diego Portales, Av. Ejercito 441, Santiago, Chile\\
}

\date{Received/accepted}

 
  \abstract
   {Hyperactive comets are a small group of comets whose activity is higher than expected. They seem to emit more water than would normally be expected given the size of their nucleus. Comet 46P/Wirtanen (hereafter, 46P) is among these objects of interest. Investigating its activity and composition evolution could provide clues about its origins and its formation region in the Solar nebulae.}
   {Given the exceptional close approach of comet 46P to the Earth in 2018, we aim to study the evolution of its activity and composition as a function of heliocentric distances before and after perihelion.}
   {We used both TRAPPIST telescopes to monitor the comet for almost a year with broad and narrow-band filters. We derived the production rates of five gaseous species (OH, NH, CN, C$_3$, and C$_2$) using a Haser model as well as the A($\theta$)f$\rho$ dust proxy parameter. The comet was also observed with the two optical high-resolution spectrographs UVES and ESPRESSO, mounted on the 8-m ESO VLT, to measure the isotopic ratios of C and N, along with the oxygen forbidden-line ratios and the NH$_2$ ortho-to-para ratios.}
   {Over nearly a year, we followed  the rise and decline of the production rates of different species as well as the dust activity of 46P on both pre- and post-perihelion. Relative abundances with respect to CN and OH along the orbit of the comet show constant and symmetric abundance ratios and a typical coma composition. We determined the rotation period of the nucleus using high-cadence observations and long series of CN images on several nights. We obtained a value of (9.18$\pm$0.05) hours at perihelion. Using the high-resolution spectra of 46P coma, we derived  C and N isotopic ratios of 100$\pm$20 and 150$\pm$30 as well as a green-to-red forbidden oxygen [OI] line ratio of 0.23$\pm$0.02. We measured a NH$_2$ ortho-to-para ratio of 3.31$\pm$0.03 and derived an ammonia ratio of 1.19$\pm$0.03, corresponding to a spin temperature of 27$\pm$1 K.}
   {Narrow-band observations show that comet 46P is a hyperactive comet for which 40\% of its nucleus surface is active. It has a typical composition, similar to other normal comets; however, an asymmetric behavior with respect to perihelion has been seen in its activity, which is typical of seasonal effects. Photometric measurements show no evidence for a change in the rotation period of the nucleus during this apparition. High-resolution spectra show that 46P has typical NH$_2$ ortho-to-para, [OI] lines ratios, and C and N isotopic ratios.}


\maketitle

\section{Introduction}\label{sec_intro}

46P/Wirtanen is a Jupiter-family comet (hereafter, JFC), with an orbital period of 5.5 years,  that was discovered on January 17, 1948 by Carl Wirtanen at the Lick Observatory \citep{Jeffers1948}. It has been classified as an hyperactive comet, namely, it is part of a small group of comets whose activity levels are higher than expected based on the sizes of their nuclei. It was an initial target of the Rosetta mission and many studies have been performed to determine its nucleus size, shape, rotation period, albedo, and color, as well as its gas and dust activity \citep{Farnham1998,Lamy1998,Schulz1998,Bertaux1999,Crovisier2002}. New radar observations have estimated the nucleus radius to be (0.70$\pm$0.05) km during its 2018 passage\footnote{ \url{https://wirtanen.astro.umd.edu/46P/}}. 

The comet's return to perihelion was highly anticipated, as it made an unusually close approach to Earth (0.07 au) in December 2018 that was only about 30 times the distance to the moon, and with an excellent visibility from both hemispheres. This passage allowed for observations to be undertaken around the world and in great detail, with a broad set of ground-based and space telescopes at a wide range of wavelengths. This close approach provided a rare spatial resolution of about 50 km/arcsec, which offered unique conditions to study the inner coma of a comet. Several results have been published on the characterization of 46P including measurements of the nucleus’ rotation period \citep{Farnham2021}, the detection of mini-outbursts \citep{Kelley2019,Kelley2021}, measurement of the D/H ratio \citep{Lis2019}, abundances of several organic molecules observed at near-infrared wavelengths at different heliocentric distances \citep{McKay2021,Khan2021,Roth2021}, search for icy grains in the coma \citep{Protopapa2021,Bonev2021}, and coma features from optical imaging \citep{Knight2021}. Millimeter and submillimeter observations of 46P have reported on the detection of several complex organic molecules and showed that the comet is relatively rich in methanol, but relatively depleted in CO, CS, HNC, HNCO, and HCOOH \citep{Biver2021,Roth2021b,Coulson2020,Bergman2022}. Using both TRAPPIST-North (TN) and TRAPPIST-South (TS), we collected photometric observation over almost a year. 46P was visible for many hours during the night, allowing for long observing sequences for months during its return, passing near to the Sun.

In this work, we report on the activity, composition, and rotation period of comet 46P during its 2018 perihelion passage. An introduction and historical background of comet 46P is given in Section \ref{sec_intro}. We describe the observing circumstances and data reduction process in Section~\ref{sec:data_reduction}. In Section~\ref{sec_activity}, we discuss the evolution of the gas and dust activity before and after perihelion and we compare the behavior of 46P  to previous apparitions. The relative molecular abundances and their evolution with respect to the heliocentric distances are discussed in the same section. In Section \ref{sec_rotation}, we measure the nucleus' rotation period and its variation around perihelion. In Section \ref{sec_spectra}, we present the C and N isotopic ratios and the NH$_2$ (and NH$_3$) ortho-to-para ratio (OPR) and the green-to-red [OI] lines (G/R) ratio derived from high-resolution spectra. The summary and conclusions of this work are given in Section \ref{sec_conclusion}.

\section{Observations and data analysis}
\label{sec:data_reduction}

\subsection{TRAPPIST}

We used both TN and TS telescopes \citep{Jehin2011} to  observe and follow comet 46P over almost a year. We started monitoring the comet at the beginning of August 2018 (\rh=1.88 au, pre-perihelion) until the end of March 2019 (\rh=1.70 au, post-perihelion). More than 2400 broad and narrow-band images of the comet were collected over 45 nights with TS and 40 nights with TN. 46P reached its perihelion on December 12, 2018 at 1.06 au from the Sun and only at 0.08 au from Earth. For the data reduction, we followed standard procedures using frequently updated master bias, flat, and dark frames. The sky contamination was removed and a flux calibration was performed using standard stars. In order to derive the production rates, we converted the  flux of different gaseous species (OH, NH, CN, C$_3$, and C$_2$), measured through the HB narrow band cometary filters \citep{Farnham2000}, to the column densities using heliocentric distance and heliocentric velocity-dependent fluorescence efficiencies \citep[and references therein]{Schleicher2010}. To convert the column densities into production rates, we used a Haser model \citep{Haser1957} to fit the coma profile. The model adjustment was performed at a physical distance of $10\,000$~km from the nucleus. The scale lengths and g-factors of different molecules at 1 au were scaled by r$_h^{-2}$ \citep{A'Hearn1995,Schleicher2010}. More details about the Haser model and its parameters are given in our previous works \citep{Moulane2018,Moulane2020}. We derived the water-production rate from our the OH production rates using the empirical formula Q(H$_2$O)=1.36 r$_h^{-0.5}$ Q(OH), given in \cite{Cochran1993,Schleicher1998}. We derived the Af$\rho$ parameter, a proxy for the dust production \citep{A'Hearn1984}, from the dust profiles using the HB cometary dust continuum BC, GC, and RC filters and the broadband R and I filters. We corrected the A($\theta$)f$\rho$ for the phase angle effect to obtain A(0)f$\rho$ using the phase function given in \cite{Schleicher2007}. The main uncertainties in the gas-production rates and Af$\rho$ come from the absolute flux calibration and the sky background subtraction. For the absolute flux calibration, we used an uncertainty of 5\% on the extinction coefficients in all filters as seen in our long-term observations of the standard stars. This is almost negligible compared to the sky level uncertainty at lower airmass, but it becomes significant at high airmasses. We estimated the three-sigma-level uncertainty on the sky background value and use it to computed the error on production rates due to the sky subtraction. As a result, errors given in the following sections are a quadratic combination of sky background and extinction coefficient uncertainties.

\subsection{UVES and ESPRESSO at 8-m ESO VLT}

We obtained a high-resolution spectrum of comet 46P with the Ultraviolet-Visual Echelle Spectrograph (UVES) mounted on the Unit 2 telescope (UT2) of the Very Large Telescopes (VLT) at the European Southern Observatory (ESO) on December 9, 2018 (a week after perihelion, r$_h$=1.05 au and $\Delta$=0.09 au). We used the UVES standard setting DIC\#1 346+580 covering the range 3030 to 3880 \AA~in the blue and 4760 to 6840 $\AA$ in the red. We used a 0.44$^{\prime\prime}$ wide slit, providing a resolving power of R$\sim$80000. We took two spectra at UT 0:59:52 (exposure time of 2300 s) and UT 1:19:54 (exposure time of 2300 s) as a start time.

The ESO UVES pipeline \citep{Dekker2000} was used to reduce the spectra in the extended object mode, in which the spatial information is kept. The spectra were corrected for  extinction and flux calibrated using the UVES master response curve provided by ESO. One-dimensional spectra were then extracted by averaging the 2D spectra with simultaneous cosmic ray rejection and then corrected for the Doppler shift due to the velocity of the comet with respect to the Earth. More details about the UVES data reduction are given in the UVES manual\footnote{\url{ftp://ftp.eso.org/pub/dfs/pipelines/uves/uves-pipeline-manual-22.17.pdf}}. The dust-reflected sunlight was finally removed using a reference solar spectrum,  BASS2000\footnote{\url{http://bass2000.obspm.fr/solar_spect.php}}. A more detailed description of the steps for the UVES data reduction and the solar spectrum subtraction is given in \cite{Manfroid2009} and references therein.

 A set of six spectra were also obtained with the new Echelle SPectrograph for Rocky Exoplanets and Stable Spectroscopic Observations (ESPRESSO) at the VLT \citep{Pepe2021} over two nights on December 9 and 10, 2018. Exposure times ranged from 5400~s for the first spectrum to 6300~s for the following five spectra. We used the ultra-high-resolution mode of the instrument, providing a spectral resolution of 190000 between 3800 and 7880 $\AA$. In this mode, the spectrograph is fed by two 0.5\arcsec-size fibers: one centered on the object and another 7\arcsec\ away usually used for simultaneous observation of the sky or a wavelength drift reference. However, in the case of the very extended coma of 46P, both fibers contained comet signal and provided an opportunity to probe different parts of the coma. Observations were executed in visitor mode, with good seeing (typically less than 1 arcsec) and the data were reduced using the ESPRESSO pipeline. which is publicly available from the ESO pipeline repository\footnote{\url{http://www.eso.org/sci/software/pipelines/espresso/espresso-pipe-recipes.html}}.

\begin{figure*}[h!]
 \sidecaption
        \includegraphics[width=1.4\columnwidth]{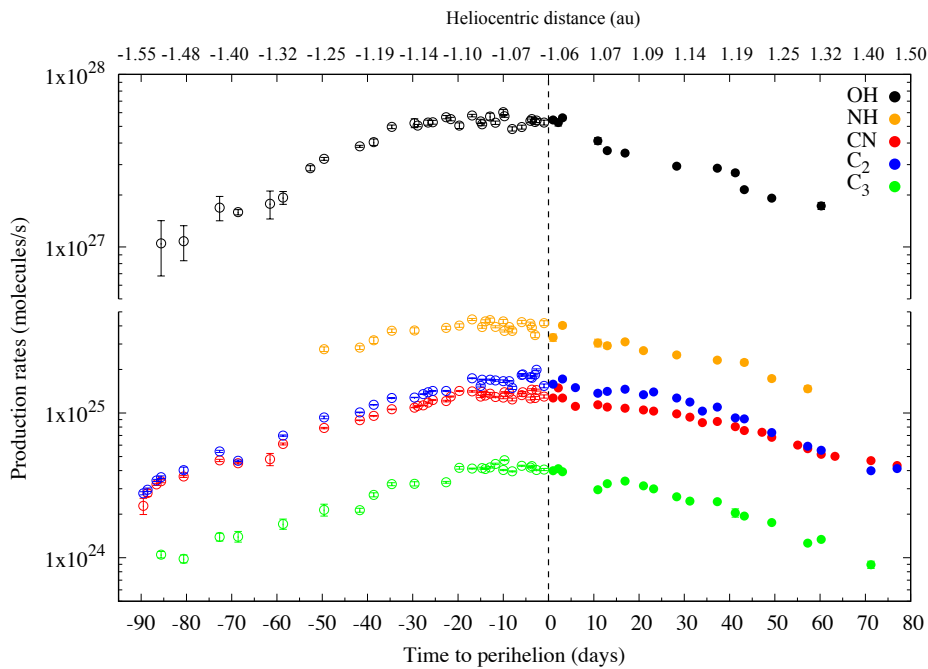}
        \caption{OH, NH, CN, C$_2$, and C$_3$ logarithmic production rates of comet 46P as a function of the heliocentric distance and of time to perihelion. The vertical dashed line indicates the perihelion at 1.06 au on December 12, 2018. Open circles indicate pre-perihelion observations, filled symbols are used for post-perihelion observations. Negative signs in x-axis represents pre-perihelion data. The values of the production rates and their relative uncertainties are given in Table \ref{tab:46P}.} 
        \label{fig:gas_rate_46P}
\end{figure*}

\section{Activity and composition}
\label{sec_activity}

\subsection{Gas-production rates and ratios}

\begin{figure}[h!]
        \centering      \includegraphics[width=0.99\columnwidth]{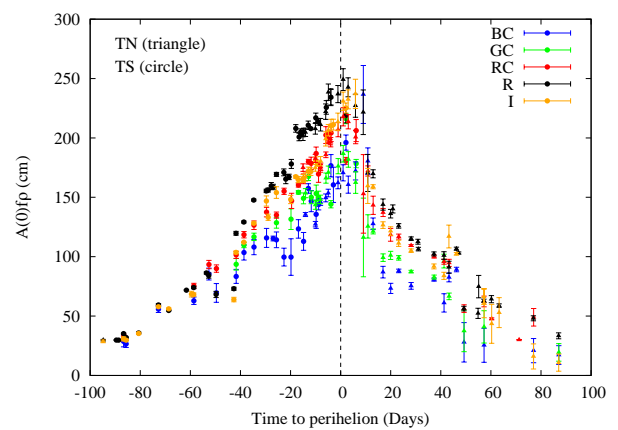}
        \caption{A(0)f$\rho$ parameter from the broad and narrow-band filters as a function of time to perihelion in days. Triangle symbols indicate data acquired with TN. Filled circles indicate TS.} 
        \label{fig:afrho_BVRI-46P}
\end{figure}

We followed the evolution of the production rates of different gaseous species over a year. The derived production rates for each species and the A(0)f$\rho$ parameter, as well as  their uncertainties are given in Table \ref{tab:46P}. Figure \ref{fig:gas_rate_46P} shows the logarithmic evolution of OH, NH, CN, C$_2$, and C$_3$ production rates as a function of the distance to the Sun and time to perihelion. The gas activity of the comet was increasing slightly as the comet was approaching the Sun and decreased rapidly after perihelion, and more sharply for NH, OH and the dust. This fall-off could be due to the orientation of the active source regions on the surface of the comet toward the Earth before and after perihelion. The same behavior has been observed in the Lowell Discovery Telescope data \citep{Knight2021}. This explanation is supported by a study of the morphological features seen in 46P coma during this passage, reported by \cite{Knight2021}. They show that 46P had large changes in both the sub-solar and sub-Earth latitudes during the apparition as well as changes in the pole orientation of the nucleus, assuming that there is no significant non-principal-axis rotation considered. It is also possible that the ice material present under the surface across the nucleus of the comet has different sublimation efficiency  at the same heliocentric distances pre- versus post-perihelion. As mentioned above, we used narrow and broadband dust filters to estimate the dust production using the Af$\rho$ parameter introduced by \cite{A'Hearn1984} (see Figure \ref{fig:afrho_BVRI-46P}). The behavior of the comet dust activity is similar to the gas on both sides of perihelion. This kind of asymmetry has been seen in many LPCs and JFCs.

\subsection{Molecular abundances and dust-to-gas ratio}

\begin{figure*}[h!]
\centering      \includegraphics[scale=0.62]{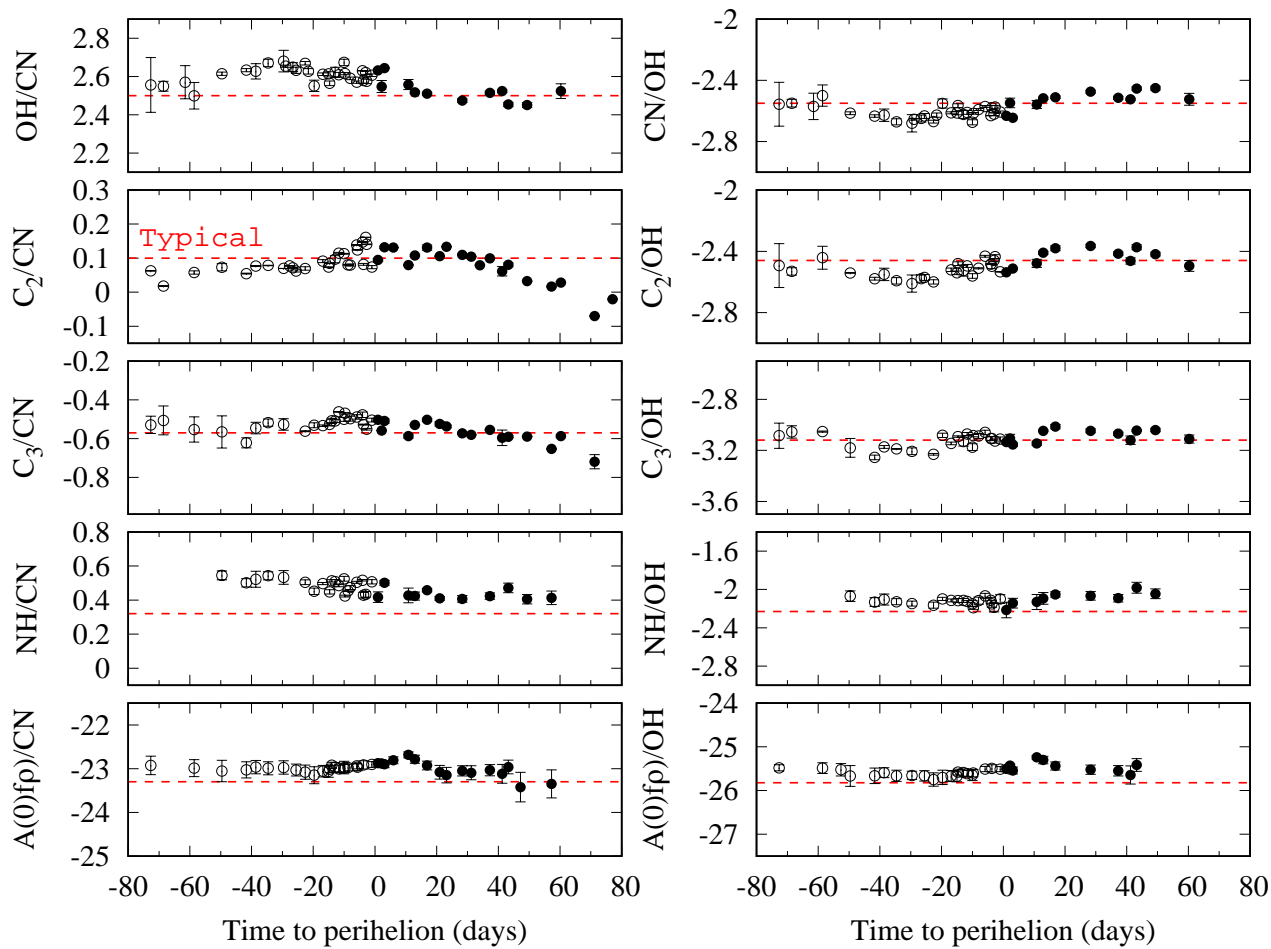}
        \caption{Logarithm of the production rate and dust-to-gas ratios  of comet 46P with respect to CN (Log10[Q(xx)/Q(CN)]) and to OH (Log10[Q(xx)/Q(OH)]) as a function of time to perihelion. Pre-perihelion values are represented with open circles and post-perihelion values with filled circles. The dust-to-gas ratio is expressed in cm second per molecule. The horizontal dashed line indicates the mean ratios of typical comets given in \cite{Schleicher2008}. } 
        \label{fig:ratios-46P}
\end{figure*}

\begin{table*}[h!]
        \begin{center}
                \caption{Mean relative molecular abundances of comet 46P for the perihelion 1991 to 2018 compared to typical and depleted comets.}
                        \label{tab:46P:abundance}
                        \resizebox{0.95\textwidth}{!}{%
                        \begin{tabular}{lccccc}
                                \hline  
                                \hline
                                &\multicolumn{5}{c}{Log production rate ratio}\\
                                \cline{2-6}
                                &1991  &1997 &  2018  & Typical comets & Depleted comets \\
                                & \citep{A'Hearn1995} &\citep{Farnham1998}  & This work & \multicolumn{2}{c}{\citep{Schleicher2008}} \\
                                \hline
                                C$_2$/CN          &-0.01 &0.10$\pm$0.10    &0.08$\pm$0.04         &  0.10$\pm$0.10 &-0.61$\pm$0.35  \\
                                C$_3$/CN        &-1.05 &-1.05$\pm$0.10   &-0.54$\pm$0.08        & -0.57$\pm$0.11 &-1.49$\pm$0.14  \\
                                CN/OH             &-2.52 &-2.50$\pm$0.12   &-2.57$\pm$0.06        & -2.55$\pm$0.18 &-2.69$\pm$0.14  \\
                                C$_2$/OH          &-2.53 &-2.36$\pm$0.30   &-2.49$\pm$0.06        & -2.46$\pm$0.20 &-3.30$\pm$0.35  \\
                                C$_3$/OH          &-3.57 &-3.49$\pm$0.12   &-3.10$\pm$0.08        & -3.12$\pm$0.29 &-4.18$\pm$0.28  \\
                                NH/OH             &-2.13 &-2.36$\pm$0.30   &-2.10$\pm$0.06        & -2.23$\pm$0.27 &-2.48$\pm$0.34  \\
                                A(0)f$\rho$/CN    &-23.36 &-23.55$\pm$0.38   &-23.05$\pm$0.20   & -23.30$\pm$0.32&-22.61$\pm$0.15 \\
                                A(0)f$\rho$/OH    &-25.88 &-25.92$\pm$0.21   &-25.62$\pm$0.25   & -25.82$\pm$0.40&-25.30$\pm$0.29 \\
                                \hline  
                                \hline
                \end{tabular}}
                \tablefoot{The Af$\rho$/Q values for typical and depleted comets are from \cite{A'Hearn1995}. We note that the difference in C$_3$ abundance in 2018 is due to the use of an updated dust continuum removal factor for the new C$_3$ filter, which results in an increase of the C$_3$ flux of about 2.1 times compared to the previous factors used in \cite{A'Hearn1995} and \cite{Schleicher1998}. More details are given in \cite{Farnham2000} and \cite{Schleicher2002}.}
        \end{center}
\end{table*}

Using the production rates above, we derived the ratios with respect to CN and to OH as well as the dust-to-gas ratios. Figure \ref{fig:ratios-46P} shows the evolution of the logarithm of the production rate ratios with respect to CN and to OH as well as ratios of A(0)f$\rho$-to-Q(OH) and A(0)f$\rho$-to-Q(CN) with time to perihelion. The C$_2$/CN ratio decreases when the comet gets far away from the Sun (40 days post-perihelion). The abundances ratios did  not show any significant variation on both sides of perihelion. This implies that 46P has an homogeneous composition along its orbit around the Sun. Comet 46P has been observed with many ground-based telescopes over the previous passages using narrow-band photometry, since it was an important target selected initially for the Rosetta mission in 1994. Table \ref{tab:46P:abundance} summarizes the mean relative molecular abundances in 46P measured in 2018 (this work), in the 1991 apparition \cite{A'Hearn1995}, and in the 1997 apparition \citep{Farnham1998} compared to the database of comets given in \citep{Schleicher2008}. The measurements in the 1991, 1997 and 2018 apparitions show that 46P has a "typical" coma composition and a normal dust-to-gas ratio.  Given the uncertainties on the ratios, no significant change is observed in the molecular abundances of 46P over the last four orbits. This shows that even an hyperactive comet, while outgassing a lot, does not change its main surface composition properties after several passages close to the Sun.

\subsection{Parent and daughter molecules}

\begin{table*}[h!]
        \caption{Abundances ratios derived from the optical and infrared data of comet 46P in its 2018 apparition.}
        \label{tab:infrared-optical}
        \resizebox{\textwidth}{!}{%
                \begin{tabular}{lcccccccccccl}
                        \hline
                        \hline
                        UT Date   & $r_h$ & $\bigtriangleup$  &  \multicolumn{7}{c}{Abundances (\%)}    & Reference \\
                        &(au) & (au)  & CN/OH & HCN/H$_2$O & C$_2$/OH & C$_2$H$_2$/H$_2$O & C$_2$H$_6$/H$_2$O & NH/OH& NH$_3$/H$_2$O & \\
                        \hline
                        2018 Dec 16 &1.06&0.07 &0.22$\pm$0.01&-&0.30$\pm$0.02&-&-&0.72$\pm$0.05&-& This work\\
                    2018 Dec 17 &1.07&0.07 &-&0.20$\pm$0.01&-&0.07$\pm$0.01&0.75$\pm$0.08&- &0.66$\pm$0.15&\cite{Bonev2021}\\
                        2018 Dec 18 &1.07&0.07 &&&&&0.75$\pm$0.08&&&\cite{Roth2021}\\
                        2018 Dec 21 &1.07&0.08 &-&0.21$\pm$0.01&-&0.08$\pm$0.01&0.71$\pm$0.09&-&0.50$\pm$0.06& \cite{Khan2021}\\
                        2018 Dec 23 &1.07&0.09 &0.22$\pm$0.01&-&0.33$\pm$0.02&-&-&0.74$\pm$0.08&-&This work\\
                        \multicolumn{11}{c}{}\\
                        2019 Jan 10 &1.12&0.17 &0.33$\pm$0.01&-&0.43$\pm$0.02&-&-&0.85$\pm$0.05&-&This work\\
                    2019 Jan 11 &1.13&0.18 &&&&&0.68$\pm$0.06&&& \cite{McKay2021}\\
                        \hline
                        \hline
        \end{tabular}}
\end{table*}

In order to investigate the origin of the radicals observed in 46P's coma, we compared our daughter species abundances with parent molecular abundances obtained at Infrared wavelengths. Table \ref{tab:infrared-optical} summarizes the abundances for some radicals observed in the optical compared to their possible parent abundances observed at NIR wavelengths \citep{Roth2021,Bonev2021,McKay2021,Khan2021}, measured during the same period of time. We compared these abundances at two epochs, perihelion (r$_h$=1.06 au) and post-perihelion (r$_h$=1.12), when the infrared observations are available.  Our CN/OH ratios are in agreement with the HCN/H$_2$O ratios derived directly from IR observations around the same dates, namely, Dec. 16-23, 2018 (see Table \ref{tab:infrared-optical}). In addition, the HCN production rates (1.06$\pm$0.03)$\times$10$^{25}$ molec/s obtained from millimeter \citep{Biver2021} and (1.60$\pm$0.10)$\times$10$^{25}$ molec/s from sub-millimeter observations \citep{Bergman2022} show an agreement with our Q(CN) values obtained during the same nights of Dec. 15 and 20. This indicates that the observed CN abundances are consistent with HCN being the dominant parent. The C$_2$ abundance is much higher than C$_2$H$_2$ which suggests that C$_2$ could be dissociated from both C$_2$H$_2$ and C$_2$H$_6$. It has been shown that C$_2$ is still linked to both molecules even at large heliocentric distances $\geq$ 3 au in comet Hale-Bopp \citep{Helbert2005}. We looked for a C$_3$ parent but none has been identified in 46P at infrared wavelengths pointing to a C$_3$ production coming from chemical reactions in the coma. Based on direct detections of NH$_3$ lines, the mixing ratio of NH$_3$/H$_2$O $\sim$0.50 in 46P (See Table \ref{tab:infrared-optical}) is lower than the NH/OH $\sim$ 0.72 obtained with TRAPPIST. This result suggests that NH is not  dissociated directly from NH$_3$, but obviously via NH$_2$ which has been demonstrated for several comets \citep{Shinnaka2011,Shinnaka2016}.

\subsection{Water-production rates and active area}

\begin{figure}[h!]
\centering      \includegraphics[width=0.99\columnwidth]{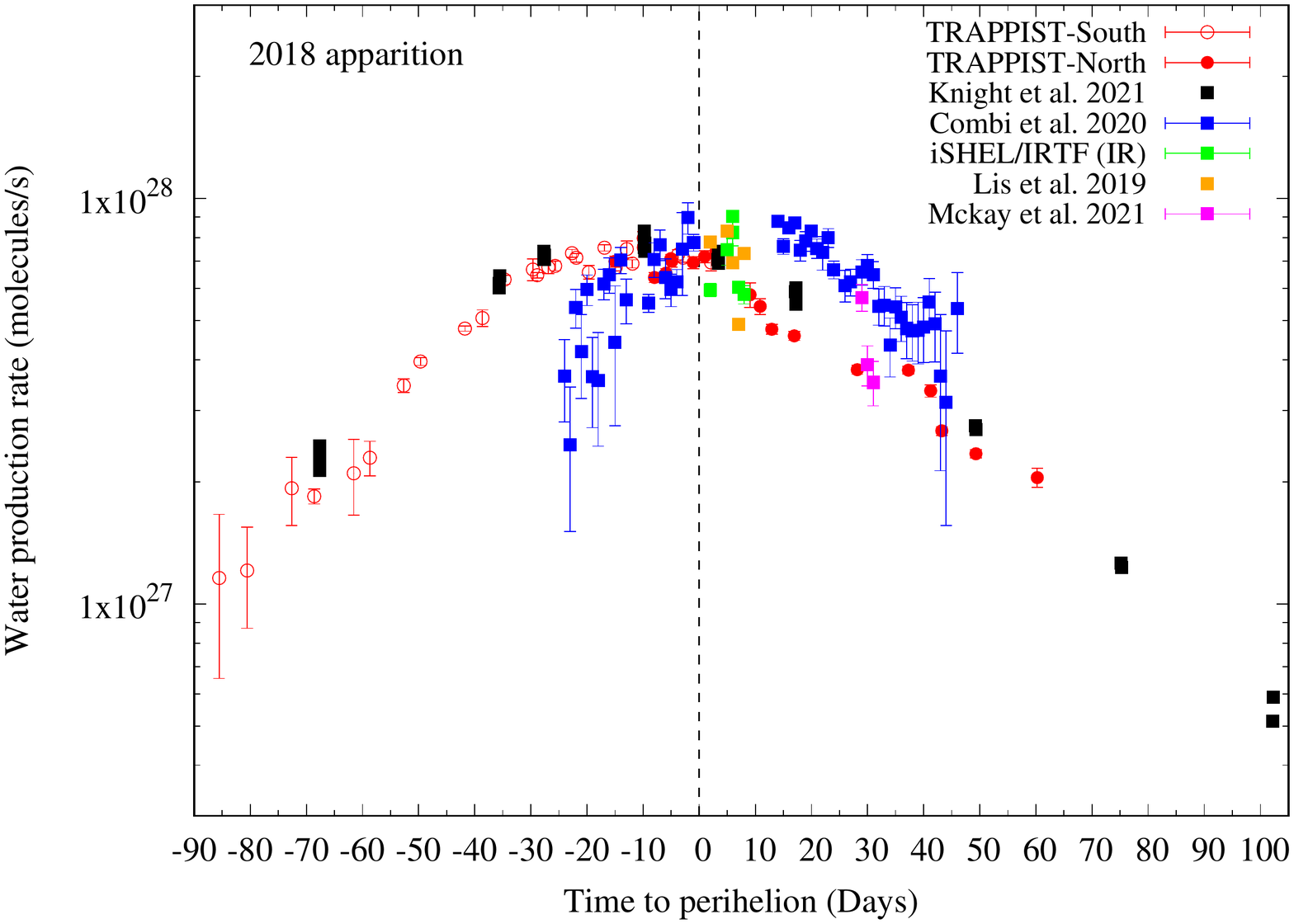}
        \caption{Water-production rate of comet 46P as a function of time from perihelion during the 2018 apparition. Our measurements and those from \cite{Knight2021} were derived from the OH production rates. \cite{Combi2020} measurements were derived from H Lyman-$\alpha$ spectra using the SWAN instrument on SOHO. iSHELL/IRTF and \citep{McKay2021} data were derived from NIR spectra. \cite{Lis2019} data were derived from far-infrared spectra using the GREAT spectrometer aboard the Stratospheric Observatory for Infrared Astronomy (SOFIA).
        } 
        \label{46P:rate_H2O_2018}
\end{figure}

\begin{figure}[h!]
\centering      \includegraphics[width=0.99\columnwidth]{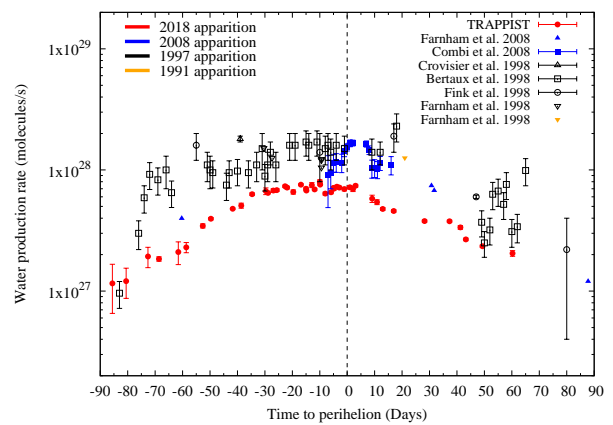}
        \caption{Water-production rates of comet 46P as a function of time from perihelion during 1991 (orange), 1997 (black), 2008 (blue), and 2018 (in red) apparitions. \cite{Fink1998} derived the water-production rate from the [OI]($^1$D) forbidden line doublet emission in the optical,  \cite{Bertaux1999} values were derived from the H Ly-$\alpha$ emission observed by the SWAN instrument on board SOHO and \cite{Crovisier2002} values were derived from the OH emission observed at the radio wavelengths.} 
                \label{46P:rate_H2O}
\end{figure}

We derived the water-production rate from OH using an empirical relationship (see Section \ref{sec:data_reduction}) proposed by \cite{Cochran1993} and \cite{Schleicher1998}. Figure \ref{46P:rate_H2O_2018} shows a comparison of our water-production rates for the 2018 apparition with other measurements derived at different wavelengths using various techniques. First, we have a very good agreement between the TRAPPIST data (red circles) and the Lowell Observatory data (black squares) on both sides of perihelion, with both using the same technique based on narrow-band filters and the same model to derive the water-production rates. We also have a good agreement with H$_2$O measured directly from the NIR spectra from iSHELL at the NASA-IRTF and NIRSPEC-2 at Keck \citep{McKay2021,Roth2021,Khan2021,Bonev2021}. The data from \cite{Combi2020} who derived the water-production rates from the hydrogen Lyman-$\alpha$ emission observed by the SWAN instrument on board  SOHO show an offset on both sides of perihelion. This disagreement is probably the result of using different techniques and models. It is also tricky to compare these results as  SWAN/SOHO instrument threshold only allows for   Lyman-$\alpha$ emission to be obtained within a small range of heliocentric distances. SWAN's values have rather large error-bars and dispersion among their measurements during this passage and the same for the previous ones (see Figure \ref{46P:rate_H2O}). We have seen this systematic difference also for comet 21P/Giacobini-Zinner \citep{Moulane2020}.

As 46P was the original target selected for Rosetta mission, many observations were performed in the previous passages in order to characterize the comet. Here, we compare the water-production rates obtained in the previous passages 1991 and 1997 with those obtained in 2008 and 2018 (see Figure \ref{46P:rate_H2O}). Besides the offset between narrow-band and spectroscopic data, the comet's activity was similar in 1991 and 1997 but 30\% lower in 2008-2018 apparitions. We did not include the 2002 observations in this comparison as they were dominated by large outbursts\footnote{\url{http://www.aerith.net/comet/catalog/0046P/2002.html}} after perihelion \citep{Combi2020}. The maximum water-production rate in 2018 was measured ten days pre-perihelion with Q(H$_2$O)=(8.00$\pm$0.30)$\times$10$^{27}$ molec/s, while it was (7.40$\pm$0.20)$\times$10$^{27}$ molec/s a month after perihelion in 2008,  
(1.23$\pm$0.10)$\times$10$^{28}$ molec/s a week before perihelion in 1997, 
and 1.25$\times$10$^{28}$ molec/s three weeks after perihelion in 1991 \citep{Farnham1998,Fink1998,Bertaux1999,Crovisier2002,Kobayashi2010}. There is a decrease  in the gas-production rates by a factor of two over the last two decades (from 1997–2018), taking into account the uncertainties and systematic offset between different techniques. Our results are similar to those reported by others who found a similar decrease rate of the water-production rate over several previous apparitions \citep{Combi2020,Knight2021}.

Comet 46P is considered a hyperactive comet as it is emitting more water than it would be expected from its nucleus radius. Based on the standard water vaporization model \citep{cowan1979}\footnote{\url{https://pdssbn.astro.umd.edu/tools/ma-evap/index.shtml}}, we estimate the evolution of the active area of 46P during our observational campaign in 2018 by modeling the water-production rate as a function of the heliocentric distances. We assumed a bond albedo of 5$\%$, 100$\%$ infrared emissivity, and the rotational pole pointed at the Sun \cite[see, e.g.,][]{hearn1989,mckay2018,mckay2019}. We found that the active area of 46P varied from $\sim$0.9~km$^{2}$ at 1.52~au pre-perihelion, reached a maximum of $\sim$2.5~km$^{2}$ at 1.06~au during the perihelion passage, and decreased to $\sim$0.7~km$^{2}$ at 1.53~au post-perihelion. Based on a radius of $\sim$0.7~km, the active fraction of the nucleus was $\sim$14\%, $\sim$40\% and $\sim$11\%, respectively. These values are much larger than the active fraction of less than 3\% derived for the majority of JFCs that have radius measurements \citep{A'Hearn1995}. Since the comet had been observed in its previous passages, we compared the active fraction of its nucleus at perihelion in the past. Based on the water-production rates derived from the empirical conversion of Haser OH production rates \citep{Schleicher1998,Farnham2007}, we estimated the active fractions of its nucleus to be $\sim$72\% at r$_h$=1.11 au in 1991, $\sim$65\% at r$_h$=1.07 au in 1997, $\sim$37\% at r$_h$=1.05 au in 2008 and $\sim$40\% at r$_h$=1.06 au in 2018. It is clear that there is a decrease in the active fraction of the nucleus with time. Comet 103P/Hartley 2 was the first known "hyperactive" comet and it was visited by the Deep Impact eXtended Investigation mission, demonstrating that this hyperactivity was produced by sublimation of icy grains in the coma \citep{AHearn2011,Protopapa2014}. Several authors reported on the presence of an extended source of water ice sublimation in 46P coma in the form of small grains or large chunks. \cite{Bonev2021} provided indirect arguments for the presence of an extended source of water vapor in the coma of 46P at NIR spectra, but they did not provide information on the properties of the ice that produces the water vapor, such as particle size and ice-to-dust ratio, while \cite{Protopapa2021} and Kareta et al. (private communication) did not detect any water ice absorption features in their NIR spectra.

\section{Rotation period and CN coma morphology}
\label{sec_rotation} 

\begin{figure}[h!]
        \includegraphics[width=0.97\columnwidth]{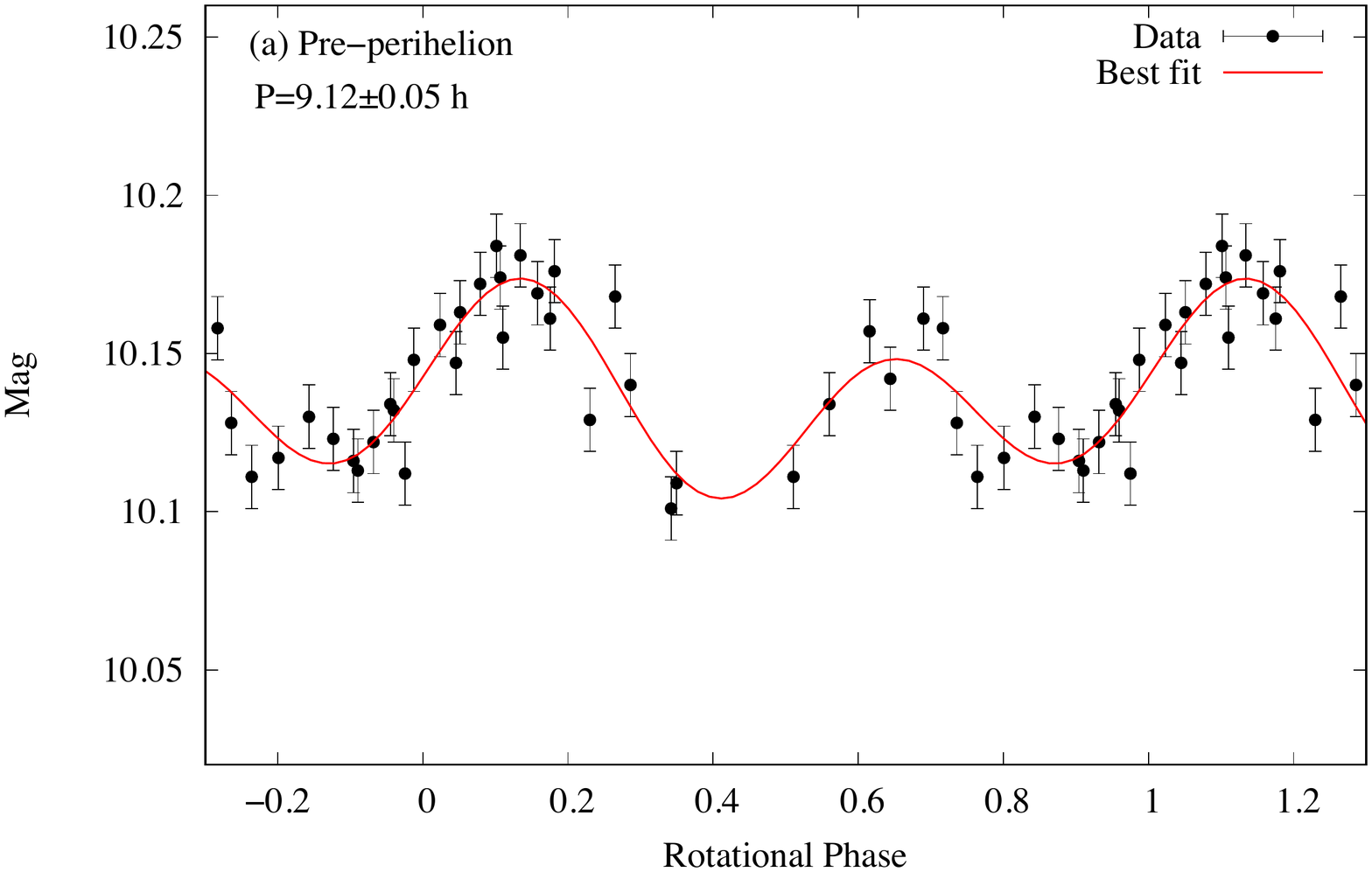}\\
        \includegraphics[width=0.97\columnwidth]{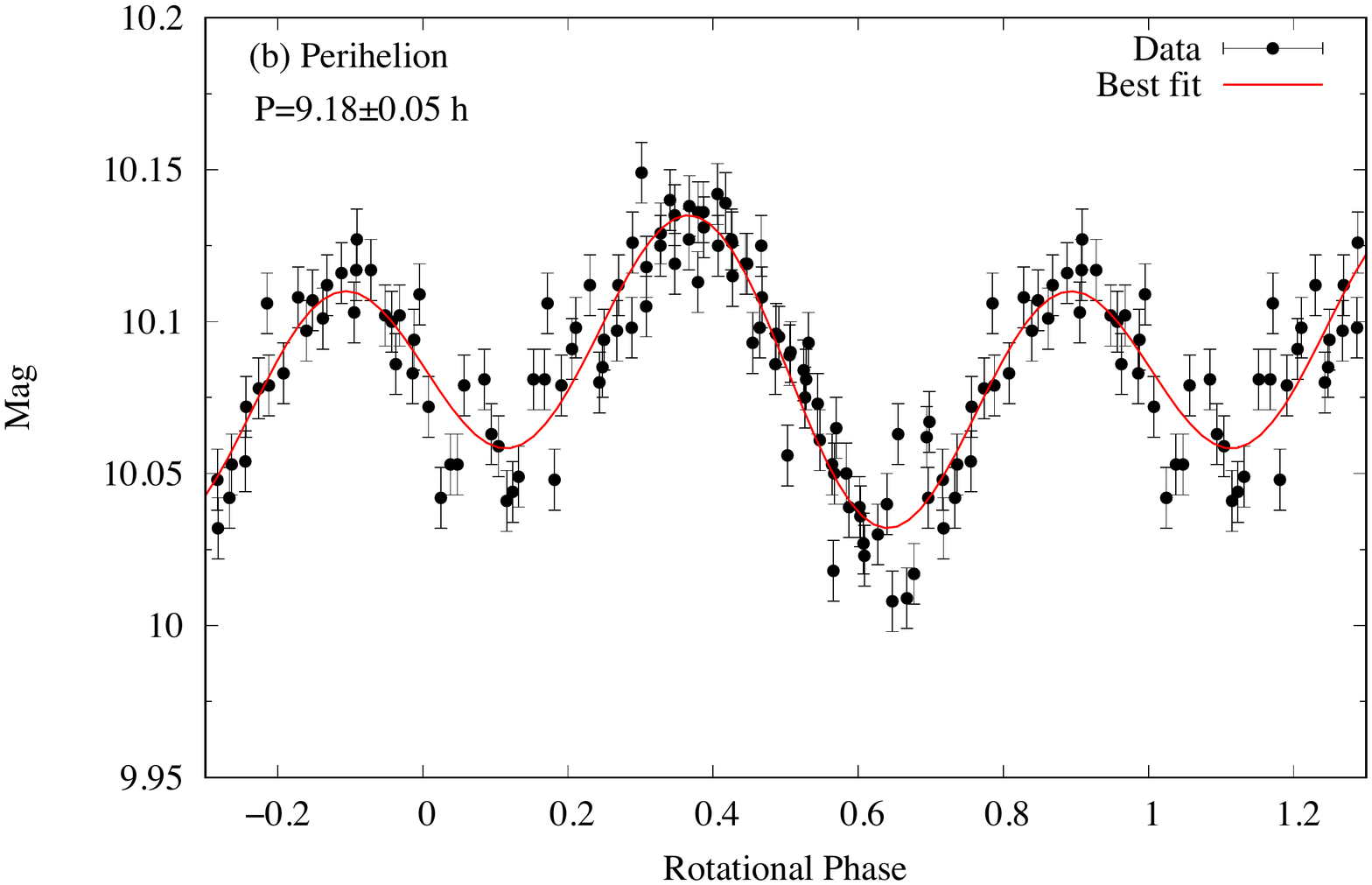} \\
        \includegraphics[width=0.97\columnwidth]{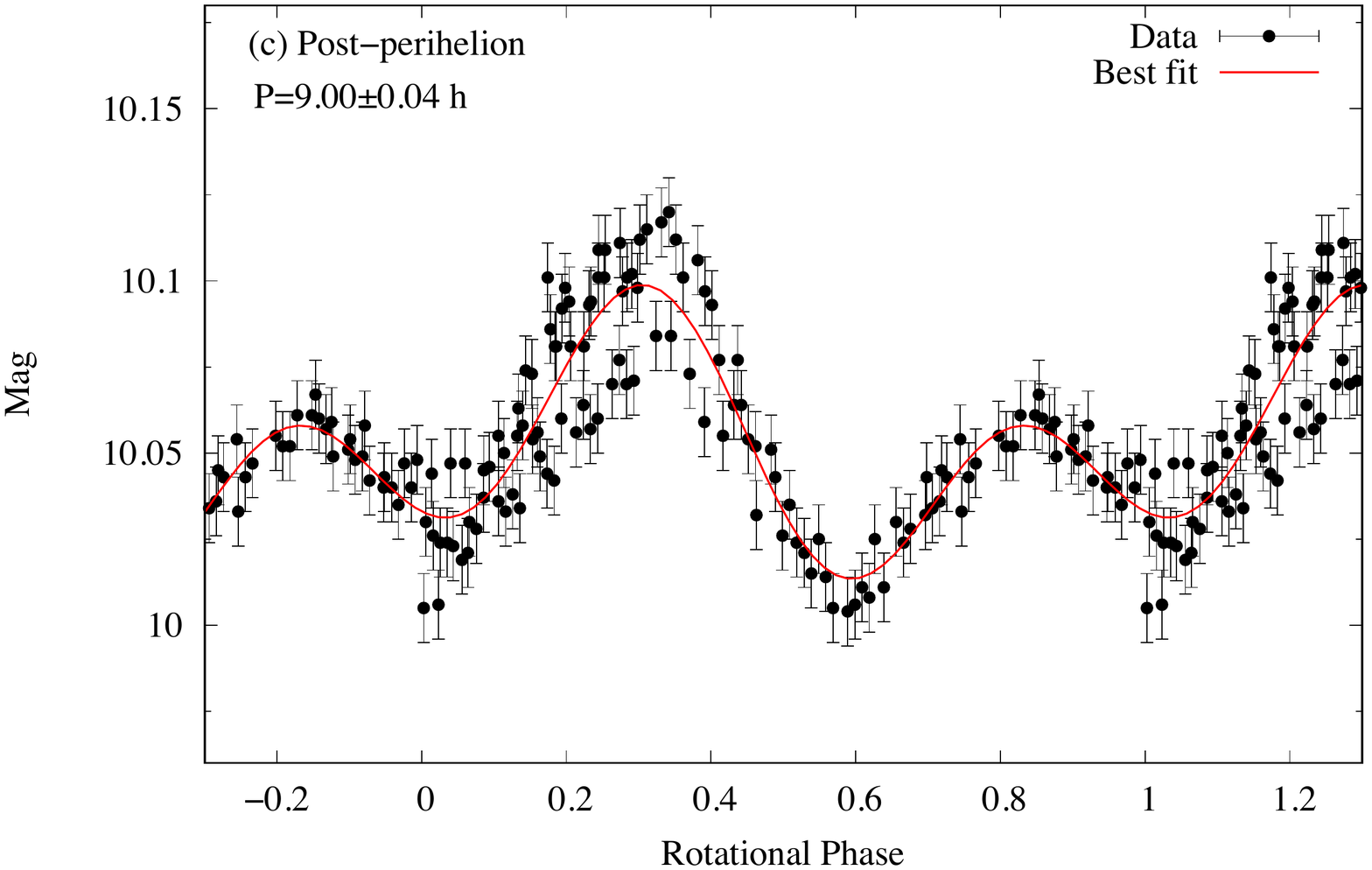}\\
        \caption{Phased CN light curves for the nucleus rotation period of comet 46P at different epochs. Time origin for the phase is JD 2458440 on Nov. 17, 2018. Data combined for (a) Nov. 23, 28 and Dec. 1, 2018 pre-perihelion, (b)  Dec. 7-10, 2018 at perihelion and for (c)  Jan. 12 and 15, 2019 post-perihelion. The red curve shows the best fit of the data.} 
        \label{CN-curves}
        \vspace{-0.2cm}
\end{figure}

We took the opportunity provided by this very close approach of 46P to investigate the rotation period of its nucleus, using long CN series of images collected during the same night and at different epochs on both sides of the perihelion. Thanks to the visibility of the comet in both hemispheres, we collected long series of CN images for many hours on several nights with both telescopes. These series were taken over 12 nights with TS and 8 nights with TN. We determined the rotation period from CN light curves. We tested different size apertures and we finally used 20$^{\prime\prime}$ which gave the largest amplitude in the light-curve. We phased multiple nights data to construct more extensive light curves and searched for the best alignment of the overlapping segments. We derived a rotation period of 9.12$\pm$0.05 hr in late November and early December (pre-perihelion), 9.18$\pm$0.05 hr between December 7 and 10 (perihelion time) and 9.00$\pm$0.04 hr in mid January 2019 (post-perihelion). These results are in agreement (within the error bars) with those derived by \cite{Farnham2021} from the CN coma morphology and photometric series (see Table \ref{rotation-period_46P}). Given the uncertainties on the measurements, we did not detect any significant change in the rotation period of the nucleus on both sides of the perihelion. Figure \ref{CN-curves} shows the CN light curves at different epochs, showing two asymmetric maxima and minima,  which could be due to two active areas on the surface of the nucleus. 

\begin{table}[h!]
\begin{center}
\caption{Rotation period of 46P nucleus at different epochs.}
\label{rotation-period_46P}
\resizebox{0.48\textwidth}{!}{%
\begin{tabular}{llcc}
\hline  
\hline
Epoch & Date UT    & \multicolumn{2}{c}{Rotation period (hour)}   \\
      &          & This work & \cite{Farnham2021}\\
\hline 
Pre-perihelion  &  Nov 23-28, 2018   & 9.12$\pm$0.05 & 9.03$\pm$0.04 \\
At perihelion   &  Dec 7-10, 2018    & 9.18$\pm$0.05 & 9.14$\pm$0.02 \\
Post-perihelion &  Jan 12-15, 2019   & 9.00$\pm$0.04 & 9.01$\pm$0.01 \\
\hline
\hline
\end{tabular}}
\end{center}
\end{table}

\begin{figure}[h!]
        \centering      \includegraphics[width=0.97\columnwidth]{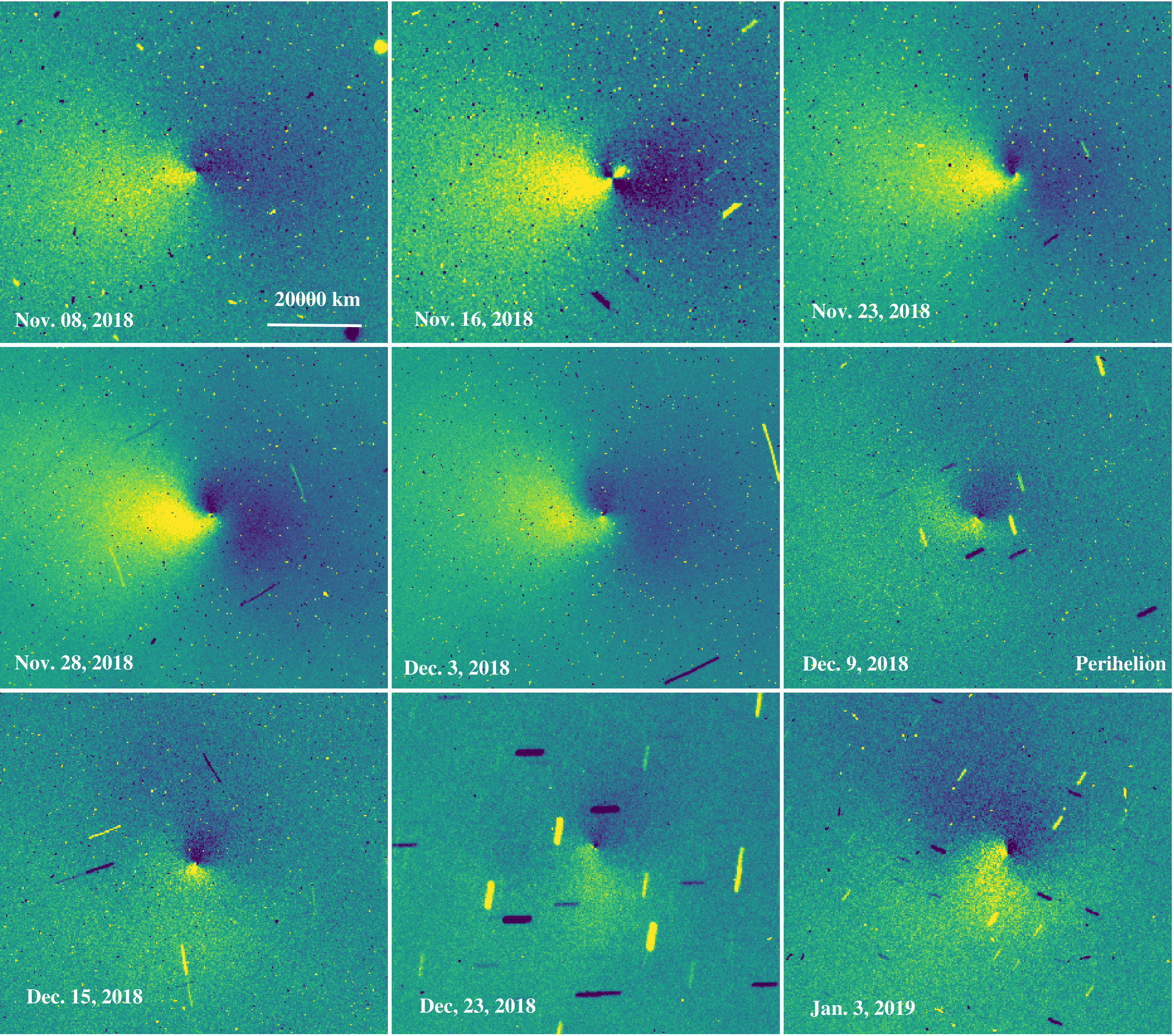}
        \caption{Evolution of the CN jets of comet 46P along its orbit around the Sun. The images are oriented north (N) up and east (E) right.}  
        \label{CN-jets}
\end{figure}

We searched for CN coma features using the rotational filter technique as described in our previous work \citep{Moulane2018}. Figure \ref{CN-jets} shows the detection and evolution of a CN jet at different epochs and on both sides of perihelion. We did not detect any clear rotation of this jet with time in our images for many hours during the same night. This could be due to an orientation of the jet towards the Earth. Around November 16, another small jet appeared in the CN images but it disappeared in the following nights. This could suggest that the nucleus might be in a non-principal-axis (NPA) rotational state, as was previously mentioned  by \cite{Samarasinha1996}. The orientation of the jet remains the same before perihelion, while it changes by $\sim$ 95 degrees counterclockwise after perihelion because of the viewing geometry of the comet with respect to Earth. \cite{Farnham2021} used the 4.3-m Lowell Discovery Telescope to investigate the CN coma morphology of 46P. They were able to detect two spiral jets in the coma. The first one appears to have been active (at varying levels) throughout most of a rotation, corresponding to the bright jet detected in our data. The second seems to turn on and off with time and it could correspond to the small jet that appears around mid November in our images. Compared to the previous apparitions, \cite{Lamy1998} derived a rotation period of the nucleus of 6.0$\pm$0.3 using data from HST, while \cite{Meech1997} found a possible rotation of 7.6 hr and a very small amplitude variation of 0.09 mag using ground-based telescope during the 1997 apparition. Based on these results, we conclude that the rotation of the 46P nucleus might have slowed down by 16\% to 45\% over the last four orbits. These changes might be due to some unusual activity in the previous apparitions. There is some evidence for potentially significant outburst activity during the 2002 apparition \citep{Combi2020} that could explain these changes.

\section{Optical high-resolution spectra}
\label{sec_spectra}

\subsection{The $^{12}$C/$^{13}$C and $^{14}$N/$^{15}$N isotopic ratios}

\begin{figure}[h!]
\vspace{-0.5cm}
\hspace{-0.2cm}  \includegraphics[width=1.0\columnwidth]{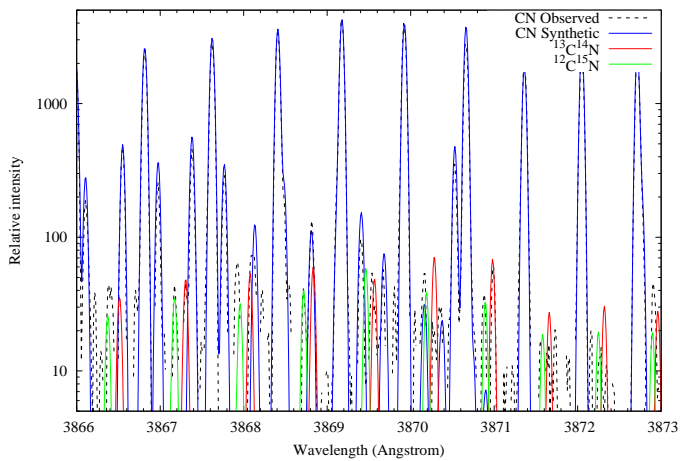}
        \centering \caption{Observed and synthetic CN isotopologues spectra of the R branch of the B-X (0, 0) violet band in comet 46P.} 
        \label{46p_CN}
\end{figure}

We obtained high-resolution observations with UVES/VLT \citep{Dekker2000} of comet 46P on December 9, 2018 (r$_h$=1.05 au and $\Delta$=0.09 au). Using the $^{12}$C$^{14}$N B-X(0,0) band, we estimated the $^{12}$C/$^{13}$C and $^{14}$N/$^{15}$N isotopic ratios of 46P. We used a CN fluorescence model to create synthetic spectra  of $^{13}$C$^{14}$N, $^{12}$C$^{15}$N, and $^{12}$C$^{14}$N. More details of the model are given in \cite{Manfroid2009}. Figure \ref{46p_CN} shows the observed CN spectrum compared to the synthetic one made under the same observing conditions. The ratios found for $^{12}$C/$^{13}$C and $^{14}$N/$^{15}$N are 100$\pm$20 and 150$\pm$30, respectively. These values are consistent with those of about 20 comets with different dynamical origins, 91.0$\pm$3.6 and 147.8$\pm$5.7 for $^{12}$C/$^{13}$C and $^{14}$N/$^{15}$N, respectively \citep{Manfroid2009,BockeleMorvan2015}. 

\subsection{The NH$_2$ and NH$_3$ ortho-to-para ratios}

\begin{figure}[h!]
        \centering      \includegraphics[width=0.99\columnwidth]{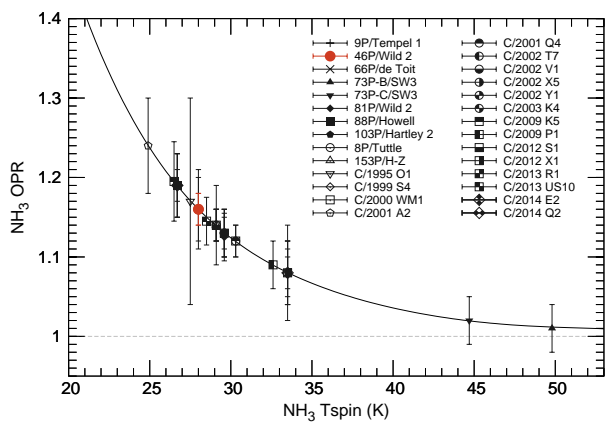}
        \caption{NH$_{3}$ $T_{\rm spin}$ in comets of various dynamical origin. The red filled circle is the value of 46P obtained with UVES at VLT (this work).}
        \label{fig:OPR_NH3_comets}
\end{figure}

From the UVES spectra, we measured the ortho-to-para (OPR, hereafter) abundance ratio of NH$_{2}$ using the three robivronic emission bands (0,7,0), (0,8,0), and (0,9,0) following the method described in  \cite{Shinnaka2011}. The derived OPRs of NH$_{2}$ and of its parent molecule NH$_{3}$ are listed for each band in Table \ref{tab_opr}, and they have average values of 3.31 $\pm$ 0.03 and 1.19 $\pm$ 0.02, respectively. A nuclear spin temperature ($T_{\rm spin}$) for ammonia of 28 $\pm$ 1 K was derived. The 46P value is consistent with typical values measured in comets (see Fig. \ref{fig:OPR_NH3_comets}), which is a possible cosmogonic indicator linked to the formation temperature of the molecule.

We would like to point out that recent laboratory experiments show that the OPR of water does not retain the memory of its formation temperature in the natal molecular cloud or in the solar nebula 4.6 Gyrs ago \citep{Hama2011, Hama2016, Hama_Watanabe2013}. The OPR of water just after its sublimation from the solid phase  is set to a statistical weight ratio (i.e., 3 for water). It is likely that this is also the case for ammonia (but the statistical weights ratio is unity for
ammonia) although no experimental studies have been reported so far. The OPRs of cometary volatiles might have been modified by some kind of ortho-to-para conversion processes in the inner coma such as water clusters, or by other catalyst activities of dust crust surfaces of the nucleus \citep{Shinnaka2016}. Thus, the observed OPRs might be diagnostic of the physico-chemical conditions in the innermost coma or beneath the surface.

\begin{table}[h!]
\caption{Derived NH$_{2}$ and NH$_{3}$ OPRs in comet 46P. Upper and lower values for each band are the results for the two spectra taken at UT 0:59:52 and UT 1:19:54 on Dec. 9, respectively.}             
\label{tab_opr}      
\centering                          
\resizebox{0.4\textwidth}{!}{%
\begin{tabular}{l c c c}        
\hline\hline                 
NH$_{2}$ band & NH$_{2}$ OPR & NH$_{3}$ OPR & $T_{\rm spin}$(K) \\    
\hline                      
   (0,7,0) & 3.40$\pm$0.09 & 1.20$\pm$0.05 & 26$^{+2}/_{-1}$ \\   
           & 3.32$\pm$0.07 & 1.16$\pm$0.04 & 28$^{+2}/_{-1}$ \\
   (0,8,0) & 3.21$\pm$0.09 & 1.11$\pm$0.05 & 31$^{+4}/_{-3}$ \\ 
           & 3.29$\pm$0.10 & 1.15$\pm$0.05 & 29$^{+3}/_{-2}$ \\
   (0,9,0) & 3.30$\pm$0.06 & 1.15$\pm$0.03 & 29$\pm$1 \\
           & 3.33$\pm$0.05 & 1.17$\pm$0.03 & 28$\pm$1 \\
\hline                      
   Average & 3.31$\pm$0.03 & 1.16$\pm$0.02 & 28$\pm$1 \\
\hline                                   
\end{tabular}}
\end{table}

\subsection{The [OI] G/R ratio}
\begin{figure}[h!]
\hspace{-0.3cm} \includegraphics[width=1.1\columnwidth]{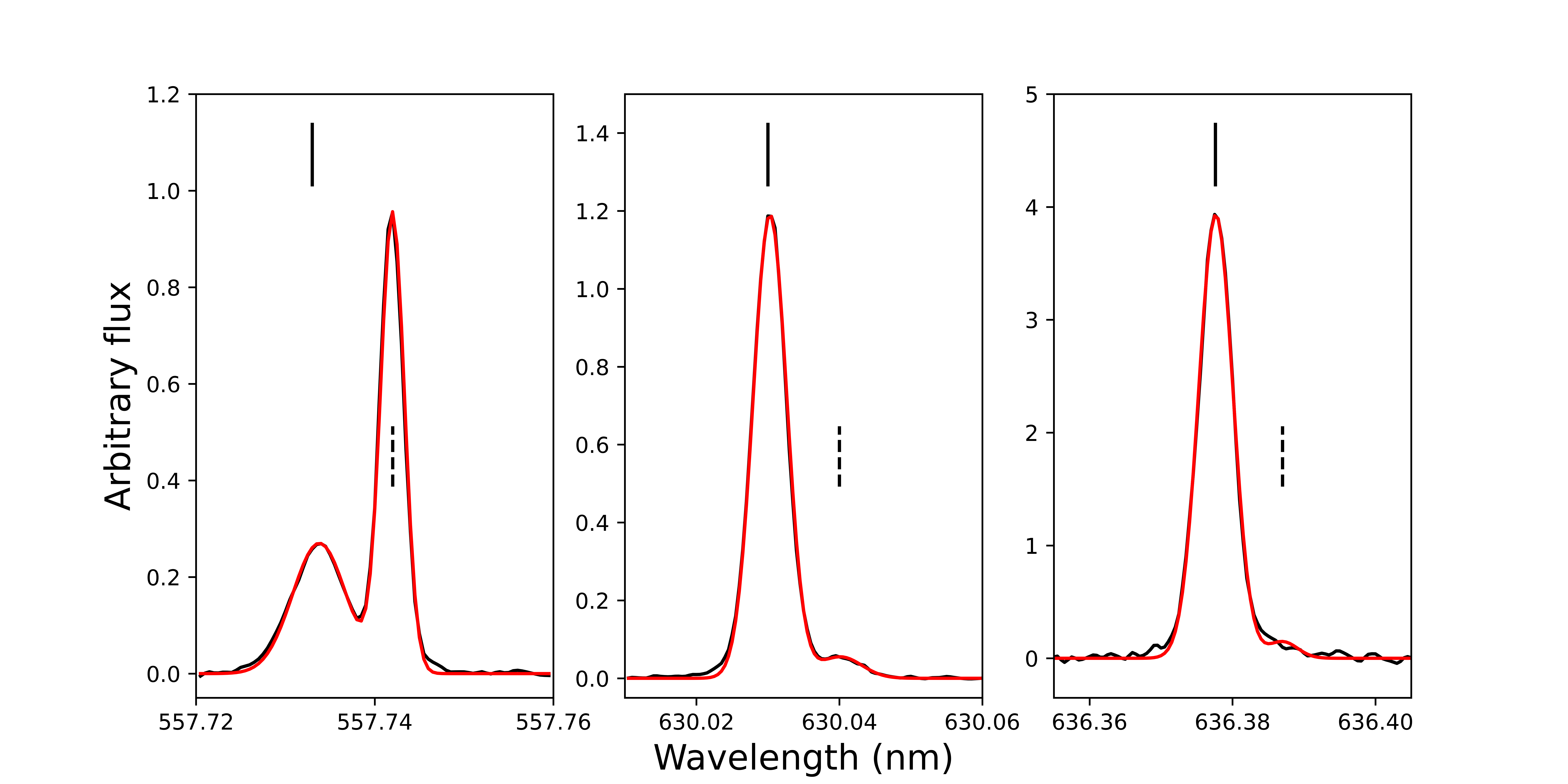}
        \caption{Three forbidden oxygen [OI] lines in comet 46P, from ESPRESSO observations, at 557.73 nm (left) for the green line, and 630.03 nm (middle) and 636.38 nm (right) for the red doublet. The cometary oxygen lines are identified by tick marks and the telluric lines marked by dashed tick. The y-axis is in arbitrary units for all three panels. The red line is a fit for one of the six spectra taken in the centered fiber.}
        \label{fig:plotGR}
\end{figure}

Using the very high-resolution ESPRESSO spectra \citep{Pepe2021}, we were able to resolve the telluric and cometary forbidden oxygen lines even with a Doppler shift as small as 5 km/s. We detected the cometary [OI] lines at 557.73, 630.03, and 636.38 nm in all six spectra (See Figure \ref{fig:plotGR}) and used them to measure the ratio between the green line (557.73 nm) and the sum of the two red lines (630.03 and 636.38 nm), commonly referred to as the G/R ratio. Due to the low geocentric velocity of the comet at the time of our observations (between -5.4 km/s and -4.5 km/s), the cometary and telluric lines were slightly blended,  even at the very high resolution of ESPRESSO. We used two Gaussian to fit both lines simultaneously. We performed the measurements for the center and the sky fiber separately for the six spectra and subsequently averaged the values for each fiber. The uncertainty of the measurement is the standard deviation of the measurements made of six spectra. We measured a G/R ratio of 0.23$\pm$0.02 in the central fiber and of 0.05$\pm$0.01 only in the sky fiber (7$\arcsec$ away). Both measurements are in agreement with what has been measured in several comets at the same nucleocentric projected distances \citep{Decock2015}. The difference of G/R ratio at different distances from the nucleus is mainly due to the collisional quenching of O($^1$S) and O($^1$D) by water molecules in the inner coma. This fits with the general behavior observed in several comets, where the G/R ratio decreases monotonically with the projected distance to the nucleus \citep{Decock2015}. The intensity of these lines can be used to derive the CO/CO$_2$/H$_2$O abundances ratios and, hence, the C/O ratio, which can provide a diagnostic of solar nebula chemistry in the comet-forming region \citep{Oberg2011}. The detail analysis and modeling of the [OI] lines are out of the scope of this paper.

\section{Summary and conclusions}
\label{sec_conclusion}

We conducted an extensive monitoring of the Jupiter Family hyperactive comet 46P on both sides of perihelion with TRAPPIST telescopes. These observations led us to an overview of the comet activity along  its journey around the Sun in 2018-2019. Using narrow and broadband filters with the TRAPPIST telescopes, we derived different quantities related to the activity of the comet such as the gas-production rates, the A(0)f$\rho$ parameter, the rotation period of the nucleus, and the morphology of the coma. Comet 46P shows an asymmetric activity with respect to perihelion, steeper decrease after perihelion for OH, NH, and dust (Af$\rho$ parameter). This might be due to the spin axis orientation or the distribution of activity on the comet's surface. The comet has exhibited a decrease in its activity from the 1997 to 2018 apparition by about 30\%. This implies a decrease of the nucleus's active fraction to $\sim$40\% which confirms its hyperactivity compared to most of JFCs. The molecular abundances relative to CN and OH remained the same as a function of the heliocentric distance, showing that 46P has a typical composition similar to what has been found in previous apparitions. These derived abundances in the optical showed a good correlation with their possible mother molecules measured in the IR (HCN, C$_2$H$_2$, and NH$_3$) and sub-millimeter (HCN), confirming the link between different molecules observed in several comets. Given the close approach of the comet, we obtained high spatial and high resolution spectra for the comet around the perihelion. Using UVES spectra, we measured typical $^{12}$C/$^{13}$C and $^{14}$N/$^{15}$N isotopic ratios of 100$\pm$20 and 150$\pm$30, and average NH$_2$ OPR of 3.31$\pm$0.03 and 1.16$\pm$0.03 for NH$_3$, with a nuclear spin temperature of 28$\pm$1 K. The ammonia OPR was found equal to 1.19$\pm$0.03,  corresponding to a spin temperature of 27$\pm$1 K. Using very-high-resolution ESPRESSO spectra, we measured a forbidden oxygen lines ratio G/R of 0.23$\pm$0.02. These measurements are in agreement with those found for several comets of different dynamical types and origins and do not reveal any peculiarity with regard to the hyperactive comet 46P.

\subsection*{Acknowledgments}
This publication makes use of data products from TRAPPIST project. TRAPPIST-South is funded by the Belgian National Fund for Scientific Research (F.R.S.-FNRS) under grant PDR T.0120.21. TRAPPIST-North is funded by the University of Liège, and performed in collaboration with Cadi Ayyad University of Marrakesh. E. Jehin and D. Hutsemékers are FNRS Senior Research Associates. J. Manfroid is Honorary Research Director of the FNRS. We thanks NASA, David Schleicher and the Lowell Observatory for the loan of a set of HB comet filters. UVES and ESPRESSO observations were collected with VLTs telescopes at the European Southern Observatory Paranal in Chile under programs 0102.C-0438(A) and 0102.C-0438(B), respectively.

\bibliographystyle{aa}
\bibliography{biblio_46P}

\onecolumn

\begin{appendix}
\section{Gas-production rates and A(0)f$\rho$ parameter of comet 46P/Wirtanen}

\begin{table*}[h!]
        \begin{center}
                \caption{Gas-production rates and A(0)f$\rho$ parameter of comet 46P/Wirtanen measured with TRAPPIST-South (TS) and -North (TN). The A(0)f$\rho$ values are computed at $10\,000$~km  from the nucleus and corrected for the phase angle effect.}
                \label{tab:46P}
                \setlength{\tabcolsep}{3.5pt}
                \resizebox{\textwidth}{!}{%
                        \begin{tabular}{lccccccccccccc}
                                \hline
                                \hline
                                UT Date & $r_h$ & $\Delta$ & $\Delta$T  & \multicolumn{5}{c}{Production rates ($\times$10$^{24}$ molecules/s) }   &  \multicolumn{4}{c}{A(0)f$\rho$ (cm)} & Tel\\
                                &(au) & (au) & (Days) & OH &NH & CN & C$_2$ & C$_3$ & BC & RC & GC & R  & \\
                                \hline
                        2018 Sep 14 &1.55 &0.63 &-89.52 &  &  &2.28$\pm$0.29 &2.79$\pm$0.14 & & & & &29.8$\pm$10.4 &TS\\
                        2018 Sep 15 &1.54 &0.62 &-88.60 &  &  &2.80$\pm$0.15 &2.95$\pm$0.16 & & & & &29.7$\pm$10.4 &TS\\
                        2018 Sep 17 &1.53 &0.60 &-86.61 &  &  &3.21$\pm$0.08 &3.42$\pm$0.12 & &28.7$\pm$14.6 &33.3$\pm$11.5 & &35.2$\pm$10.6&TS\\
                        2018 Sep 18 &1.52 &0.59 &-85.60 &1050$\pm$371 &  &3.38$\pm$0.09 &3.60$\pm$0.15 &1.05$\pm$0.05 &26.6$\pm$13.0 & & &32.0$\pm$10.4 &TS\\
                        2018 Sep 23 &1.48 &0.55 &-80.59 &1080$\pm$248 &  &3.63$\pm$0.08 &4.03$\pm$0.20 &1.00$\pm$0.06 & & & &35.8$\pm$10.8 &TS\\
                        2018 Oct 01 &1.42 &0.48 &-72.65 &1690$\pm$271 &  &4.70$\pm$0.10 &5.43$\pm$0.22 &1.39$\pm$0.08 &55.8$\pm$12.8 & & &59.4$\pm$11.1&TS\\
                        2018 Oct 05 &1.38 &0.45 &-68.58 &1590$\pm$100 &  &4.49$\pm$0.07 &4.68$\pm$0.10 &1.40$\pm$0.12 & & & &54.7$\pm$10.4 &TS\\
                        2018 Oct 12 &1.33 &0.39 &-61.54 &1780$\pm$328 &  &4.79$\pm$0.06 & & & & & &71.8$\pm$10.5 &TS\\
                        2018 Oct 15 &1.30 &0.38 &-58.63 &1930$\pm$164 &  &6.11$\pm$0.11 &6.98$\pm$0.14 &1.71$\pm$0.13 &62.9$\pm$13.1 &75.5$\pm$12.1 & &74.0$\pm$11.0 &TS\\
                        2018 Oct 21 &1.26 &0.34 &-52.62 &2860$\pm$180 &  &8.73$\pm$0.09 &10.50$\pm$0.13 &2.63$\pm$0.11 &85.4$\pm$15.0 &93.4$\pm$13.5 & &84.1$\pm$12.2 &TS\\
                        2018 Oct 24 &1.24 &0.32 &-49.60 &3240$\pm$120 &27.60$\pm$1.03 &7.87$\pm$0.12 &9.32$\pm$0.21 &2.14$\pm$0.20 &69.5$\pm$17.9 &90.0$\pm$13.1 & &68.2$\pm$12.6 &TS\\
                        2018 Nov 01 &1.19 &0.27 &-41.71 &3830$\pm$110 &28.30$\pm$0.84 &8.91$\pm$0.08 &10.10$\pm$0.10 &2.13$\pm$0.07 &83.4$\pm$15.9 &101.5$\pm$13.0 & 93.6$\pm$14.9 &119.7$\pm$11.6 &TS\\
                        2018 Nov 04 &1.18 &0.26 &-38.61 &4050$\pm$176 &31.80$\pm$1.61 &9.55$\pm$0.06 &11.40$\pm$0.09 &2.72$\pm$0.09 &103.5$\pm$16.3 &118.5$\pm$12.7 &109.8$\pm$12.2 &129.2$\pm$10.8 &TS\\
                        2018 Nov 08 &1.15 &0.23 &-34.61 &4970$\pm$119 &37.00$\pm$1.03 &10.60$\pm$0.10 &12.70$\pm$0.11 &3.22$\pm$0.07 &108.1$\pm$16.4 &126.5$\pm$13.6 &116.7$\pm$13.1 &147.7$\pm$10.8 &TS\\
                        2018 Nov 13 &1.13 &0.21 &-29.61 &5220$\pm$310 &35.20$\pm$1.60 &10.90$\pm$0.06 &12.80$\pm$0.13 &3.24$\pm$0.11 &115.8$\pm$18.0 &137.5$\pm$14.0 &  &155.4$\pm$11.4 &TS\\
                        2018 Nov 14 &1.13 &0.20 &-28.90 &5050$\pm$140 &  &11.20$\pm$0.15 & & & & & &156.1$\pm$11.7  &TS\\
                        2018 Nov 15 &1.12 &0.20 &-27.67 &5260$\pm$177 &  &11.30$\pm$0.08 &13.50$\pm$0.13 & & & & &159.6$\pm$12.9 &TS\\
                        2018 Nov 17 &1.11 &0.19 &-25.57 &5280$\pm$133 &  &12.30$\pm$0.16 &13.90$\pm$0.18 & &114.2$\pm$16.6 &135.0$\pm$12.6 &128.6$\pm$14.6  &169.3$\pm$11.7  &TS\\
                        2018 Nov 20 &1.10 &0.17 &-22.62 &5650$\pm$122 &34.70$\pm$1.03 &12.10$\pm$0.18 &14.20$\pm$0.20 &3.32$\pm$0.05 &99.7$\pm$17.5 &155.1$\pm$12.6 & &171.0$\pm$12.9  &TS\\
                        2018 Nov 21 &1.10 &0.17 &-21.58 &5510$\pm$130 &  &13.00$\pm$0.20 & & & & & &165.3$\pm$13.8 &TS\\
                        2018 Nov 23 &1.09 &0.16 &-19.70 &5050$\pm$191 &40.20$\pm$1.10 &14.20$\pm$0.19 & &4.18$\pm$0.16&99.7$\pm$20.4 &147.0$\pm$16.4 &131.6$\pm$18.7 &178.1$\pm$13.8&TS\\
                        2018 Nov 26 &1.08 &0.14 &-16.89 &5770$\pm$101 &44.30$\pm$0.90 &14.10$\pm$0.18 &17.40$\pm$0.22 &4.13$\pm$0.05 &123.4$\pm$17.8 &160.3$\pm$11.7 &154.2 $\pm$11.5 &200.8$\pm$13.2 &TS\\
                        2018 Nov 27 &1.08 &0.14 &-15.90 &  &  &13.70$\pm$0.21 & & & & & &205.0$\pm$12.6  &TS\\
                        2018 Nov 27 &1.08 &0.13 &-14.99 &5340$\pm$155 & &13.20$\pm$0.23 &15.40$\pm$0.26 &4.10$\pm$0.06&112.9$\pm$17.5&165.6$\pm$14.1 &149.1$\pm$14.6 &201.6$\pm$13.8 &TN\\
                        2018 Nov 28 &1.07 &0.13 &-14.66 &5140$\pm$102 &39.30$\pm$0.80 &14.00$\pm$0.20 &17.00$\pm$0.22 &4.16$\pm$0.06 &135.5$\pm$11.7 &175.0$\pm$13.2 &153.5$\pm$13.8 &205.3$\pm$12.9 &TS\\
                        2018 Nov 30 &1.07 &0.13 &-12.90 &5700$\pm$265 &43.80$\pm$1.50 &13.60$\pm$0.30 &17.00$\pm$0.32 &4.21$\pm$0.09 &157.5$\pm$14.0 &180.0$\pm$12.0 &167.6$\pm$13.5 &210.7$\pm$13.5 &TS\\
                        2018 Dec 01 &1.07 &0.12 &-11.73 &5250$\pm$117 &39.40$\pm$0.75 &12.90$\pm$0.26 &16.80$\pm$0.21 &4.45$\pm$0.10 &146.8$\pm$18.9 &178.6$\pm$15.2 &150.0$\pm$15.4 &207.8$\pm$10.6 &TS\\
                        2018 Dec 02 &1.06 &0.11 &-10.01 &6040$\pm$219 &42.90$\pm$1.46 &12.80$\pm$0.25 &16.60$\pm$0.35 &4.04$\pm$0.12 &127.4$\pm$12.3 &182.7$\pm$13.4 &143.6$\pm$13.7 &207.7$\pm$14.5 &TN\\
                        2018 Dec 03 &1.06 &0.11 &-09.74 &5730$\pm$110 &36.90$\pm$1.00 &13.90$\pm$0.28 & &4.72$\pm$0.08 &135.7$\pm$21.9 &186.9$\pm$15.4 &153.2$\pm$17.1 &216.9$\pm$14.0  &TS\\
                        2018 Dec 04 &1.06 &0.11 &-08.68 & &39.30$\pm$0.62 &13.80$\pm$0.32 &16.60$\pm$0.25 & &143.9$\pm$14.5 &169.6$\pm$17.0 &149.9$\pm$15.3 &213.6$\pm$12.8  &TS\\
                        2018 Dec 04 &1.06 &0.10 &-08.01 &4830$\pm$135 &36.90$\pm$1.20 &12.40$\pm$0.26 &14.90$\pm$0.40 &3.95$\pm$0.10 &146.0$\pm$12.8 & 174.4$\pm$13.3&148.2$\pm$13.6  &211.1$\pm$15.3  &TN\\
                        2018 Dec 07 &1.06 &0.10 &-05.71 &  &  &13.90$\pm$0.33 &18.50$\pm$0.22 & & &202.4$\pm$17.3 & &225.7$\pm$15.7 &TS\\
                        2018 Dec 07 &1.06 &0.10 &-05.90 &4950$\pm$131 &42.50$\pm$1.12 &12.50$\pm$0.27 &18.40$\pm$0.28 &4.32$\pm$0.11 &149.5$\pm$13.3 &193.1$\pm$13.2 &166.2$\pm$14.3 &222.1$\pm$16.5 &TN\\
                        2018 Dec 08 &1.06 &0.19 &-04.93 &5380$\pm$130 &41.40$\pm$1.14 &14.20$\pm$0.28 &17.70$\pm$0.45 &4.20$\pm$0.08 &153.6$\pm$13.2 &198.2$\pm$14.2 &171.0$\pm$15.3 &238.5$\pm$16.9 &TN\\
                        2018 Dec 09 &1.06 &0.09 &-04.00 &  &  &13.60$\pm$0.30 &17.50$\pm$0.43 &4.10$\pm$0.11 & & & &234.3$\pm$17.1  &TN\\
                        2018 Dec 09 &1.06 &0.09 &-03.69 &5050$\pm$152 &38.90$\pm$1.21 &14.50$\pm$0.40 &17.50$\pm$0.40 &4.30$\pm$0.09 &176.6$\pm$19.4 &204.2$\pm$17.3 &144.2$\pm$13.1 &234.3$\pm$16.5 &TS\\
                        2018 Dec 10 &1.06 &0.09 &-02.98 &5030$\pm$166 &34.50$\pm$1.34 &13.70$\pm$0.28 &18.40$\pm$0.41 & & & & &  &TN\\
                        2018 Dec 10 &1.06 &0.09 &-02.65 &5410$\pm$144 & &14.40$\pm$0.42 &19.90$\pm$0.50 &4.04$\pm$0.09 & 162.2$\pm$15.0&207.0$\pm$16.0 &207.0$\pm$16.0 &236.9$\pm$17.2 &TS\\
                        2018 Dec 12 &1.06 &0.08 &-01.00 &5260$\pm$173 &41.80$\pm$1.63 &13.00$\pm$0.34 &15.40$\pm$0.62 &4.07$\pm$0.12 & & & & &TN\\
                        2018 Dec 14 &1.06 &0.08 &+01.01 &5440$\pm$180 &33.30$\pm$1.89 &13.70$\pm$0.32 &15.80$\pm$0.47 & 4.00$\pm$0.08&170.6$\pm$16.8 &217.8$\pm$17.7 &186.7$\pm$17.0 & 249.1$\pm$19.2&TN\\
                        2018 Dec 15 &1.06 &0.08 &+02.17 &5260$\pm$237 & &14.90$\pm$0.41 & &4.12$\pm$0.05 &196.1$\pm$16.4 &180.7$\pm$12.9 &216.2$\pm$14.2 &219.0$\pm$16.3  &TS\\
                        2018 Dec 16 &1.06 &0.08 &+03.10 &5600$\pm$154 &40.30$\pm$1.58 &13.70$\pm$0.34&17.20$\pm$0.50&3.93$\pm$0.06 &160.7$\pm$17.2 & 213.6$\pm$16.0&181.9$\pm$16.6 & 242.6$\pm$18.0&TN\\
                        2018 Dec 17 &1.06 &0.08 &+04.02 &  &  &13.10$\pm$0.34 & & & & & & &TN\\
                        2018 Dec 17 &1.06 &0.08 &+04.02 &  &  &14.60$\pm$0.40 & & & & & & &TS\\
                        2018 Dec 18 &1.06 &0.08 &+05.95 & & &12.30$\pm$0.26 &15.00$\pm$0.38 & &172.2$\pm$17.6 &200.7$\pm$15.6 &162.4$\pm$19.1 & 227.1$\pm$19.7&TN\\
                        2018 Dec 21 &1.06 &0.08 &+08.01 & & & & & &236.6 $\pm$20.4 &152.9$\pm$20.1 &116.2$\pm$20.1 & 221.6$\pm$28.8&TN\\
                        2018 Dec 23 &1.07 &0.09 &+10.88 &4120$\pm$183 &30.50$\pm$1.88 &11.40$\pm$0.21 &13.70$\pm$0.48 &2.95$\pm$0.16 &180.5$\pm$20.1 &172.2$\pm$14.3 &125.5$\pm$19.8 & 170.5$\pm$25.6&TN\\
                        2018 Dec 25 &1.07 &0.10 &+12.95 &3620$\pm$110 &29.20$\pm$1.28 &11.20$\pm$0.23 &14.10$\pm$0.46 &3.25$\pm$0.12 &127.7$\pm$14.9 &143.3$\pm$17.7 &121.3$\pm$12.1 & 169.8$\pm$13.4&TN\\
                        2018 Dec 29 &1.08 &0.11 &+16.91 &3500$\pm$100 &31.00$\pm$0.89 &10.80$\pm$0.18 &14.60$\pm$0.26 &3.39$\pm$0.08 & 87.0$\pm$14.9 &137.2$\pm$13.4 & 99.2$\pm$13.2 & 143.9$\pm$14.7&TN\\
                        2019 Jan 02 &1.09 &0.13 &+20.98 &  &27.00$\pm$0.95  &10.50$\pm$0.25 &13.40$\pm$0.40 &3.14$\pm$0.05 &73.3$\pm$14.3 &118.4$\pm$15.4 &101.1$\pm$13.4 &140.2$\pm$13.4 &TN\\
                        2019 Jan 05 &1.10 &0.14 &+23.21 & & &10.30$\pm$0.17 &14.00$\pm$0.44 &3.00$\pm$0.06 &88.0$\pm$11.5 &116.9$\pm$12.1 &99.0$\pm$11.9 & 125.9$\pm$11.9&TN\\
                        2019 Jan 10 &1.12 &0.17 &+28.32 &2940$\pm$115 &25.50$\pm$0.90 &9.89$\pm$0.14 &12.70$\pm$0.32 &2.64$\pm$0.07 &75.4$\pm$12.8 &109.4$\pm$10.8 &87.4$\pm$11.1 &115.2$\pm$11.5  &TN\\
                        2019 Jan 13 &1.14 &0.19 &+31.22 &  &  &9.36$\pm$0.11 &12.00$\pm$0.26 &2.46$\pm$0.07 & & & &106.4$\pm$11.5  &TN\\
                        2019 Jan 15 &1.15 &0.21 &+34.01 &  &  &8.58$\pm$0.09 &10.30$\pm$0.22 & & & & &  &TN\\
                        2019 Jan 19 &1.17 &0.23 &+37.29 &3000$\pm$120 &23.20$\pm$0.85 &8.75$\pm$0.10 &11.00$\pm$0.23 &2.44$\pm$0.05 &80.7$\pm$11.3 &100.1$\pm$11.3 &82.8$\pm$11.7 &102.3$\pm$11.7&TN\\
                        2019 Jan 23 &1.19 &0.25 &+41.27 &2690$\pm$105 &22.90$\pm$1.67 &8.05$\pm$0.21 &9.27$\pm$0.20 &2.04$\pm$0.06  &61.1$\pm$14.7 & 95.8$\pm$12.1 &96.3$\pm$15.8 &101.0$\pm$13.2&TN\\
                        2019 Jan 25 &1.20 &0.27 &+43.26 &2150$\pm$110 &24.40$\pm$0.95 &7.56$\pm$0.09 &9.10$\pm$0.10 &1.94$\pm$0.07  &82.6$\pm$13.8 & 94.8$\pm$11.0 &66.6$\pm$12.8 & 91.1$\pm$15.1&TN\\
                        2019 Jan 28 &1.22 &0.28 &+46.15 & & &8.37$\pm$0.08 & & & & & &106.7$\pm$11.3&TN\\
                        2019 Jan 31 &1.24 &0.31 &+49.31 &1920$\pm$100 &17.30$\pm$1.15 &6.80$\pm$0.07 &7.33$\pm$0.13 &1.75$\pm$0.04&27.9$\pm$13.6 &56.3$\pm$10.2 &37.5$\pm$15.5 &56.3$\pm$11.9  &TN\\
                        2019 Feb 05 &1.28 &0.35 &+55.04 &  &  &6.01$\pm$0.05 & & & & & &52.3$\pm$14.3  &TN\\
                        2019 Feb 08 &1.30 &0.37 &+57.26 & &14.70$\pm$0.90 &5.67$\pm$0.06 &5.89$\pm$0.08 &1.26$\pm$0.03 &25.4$\pm$12.3 &62.3$\pm$11.5 &40.9$\pm$10.6 &62.9$\pm$11.3&TN\\
                        2019 Feb 11 &1.32 &0.39 &+60.15 &1730$\pm$95 &16.30$\pm$0.95 &6.18$\pm$0.07 &5.53$\pm$0.12 &1.34$\pm$0.05 & & & &64.2$\pm$13.0&TN\\
                        2019 Feb 14 &1.34 &0.41 &+63.25 & & &5.02$\pm$0.07 & & & & & &58.8$\pm$11.5&TN\\
                        2019 Feb 22 &1.40 &0.49 &+71.30 & & &5.69$\pm$0.10 &4.00$\pm$0.16 &0.90$\pm$0.10 & & & &&TN\\
                        2019 Feb 28 &1.45 &0.55 &+77.45 & & &4.33$\pm$0.04 &4.13$\pm$0.08 & & & & &48.1$\pm$12.1&TN\\
                        2019 Mar 10 &1.53 &0.65 &+87.10 & &  &3.16$\pm$0.05 &2.81$\pm$0.10 &0.60$\pm$0.03 &17.2$\pm$10.9 & &19.4$\pm$11.4 &33.4$\pm$12.5&TN\\
                        2019 Mar 20 &1.62 &0.77 &+97.35 &  &  &3.05$\pm$0.13 &2.00$\pm$0.13 & & & & &&TN\\
                                \hline  
                                \hline
                \end{tabular}}
        \end{center}
\end{table*}
\end{appendix}

\end{document}